\documentclass[aps,prl,reprint,groupedaddress,showpacs]{revtex4-1}

\usepackage{graphicx}
\usepackage{natbib}
\usepackage{amsmath}

\begin{document}


\title{Size Dependence of Domain Pattern Transfer in Multiferroic Heterostructures}

\author{K\'{e}vin J. A. Franke, Diego L\'{o}pez Gonz\'{a}lez, Sampo J. H\"{a}m\"{a}l\"{a}inen, and Sebastiaan van Dijken}
\email[]{sebastiaan.van.dijken@aalto.fi}
\affiliation{NanoSpin, Department of Applied Physics, Aalto University School of Science, P.O. Box 15100, FI-00076 Aalto, Finland.}


\date{\today}

\begin{abstract}
Magnetoelectric coupling in multiferroic heterostructures can produce large lateral modulations of magnetic anisotropy enabling the imprinting of ferroelectric domains into ferromagnetic films. Exchange and magnetostatic interactions within ferromagnetic films oppose the formation of such domains. Using micromagnetic simulations and a one-dimensional model, we demonstrate that competing energies lead to the breakdown of domain pattern transfer below a critical domain size. Moreover, rotation of the magnetic field results in abrupt transitions between two scaling regimes with different magnetic anisotropy. The theoretical predictions are confirmed by experiments on CoFeB/BaTiO$_3$ heterostructures.
\end{abstract}

\maketitle

Electric-field control of magnetism has recently attracted considerable interest as a low-power alternative for spintronic devices. One promising approach focuses on magnetoelectric interface coupling between a ferroelectric material and a thin ferromagnetic film in multiferroic heterostructures \cite{2007NatMa...6...21R,ADMA:ADMA201003636,0953-8984-24-33-333201}. In such structures, ferroelectric polarization reversal between two out-of-plane states has been used to alter the magnetic interface properties via electrostatic charge modulation or electronic hybridization effects \cite{2010Sci...327.1106G,2012NatMa..11..289P,2013NatMa..12..397Y}. In addition, domain pattern transfer from a ferroelectric material to a ferromagnetic film has been demonstrated for systems with varying in-plane polarization. The transfer of ferroelectric domains originates from a magnetoelectric coupling mechanism whereby the in-plane component of the ferroelectric polarization induces a local uniaxial magnetic anisotropy in the ferromagnetic material. Two mechanisms have thus far been identified, namely, exchange coupling between the canted magnetic moment of ferroelectric BiFeO$_3$ and an adjacent ferromagnetic layer \cite{2008NatMa...7..478C,2009PhRvL.103y7601L,2011PhRvL.107u7202H} and strain transfer from ferroelectric domains in BaTiO$_3$ to a magnetostrictive film \cite{ADMA:ADMA201100426,2011ITM....47.3768L,2012PhRvB..86a4408C,2012ApPhL.101z2405L}. Since the orientation of uniaxial magnetic anisotropy is linked to the direction of ferroelectric polarization, electric-field control of local magnetic switching and the writing of magnetic domain patterns are possible. Moreover, electric-field induced magnetic domain wall motion has recently been demonstrated in zero applied magnetic field \cite{2012NatSR...2E.258L}.       
  
\begin{figure}
\includegraphics{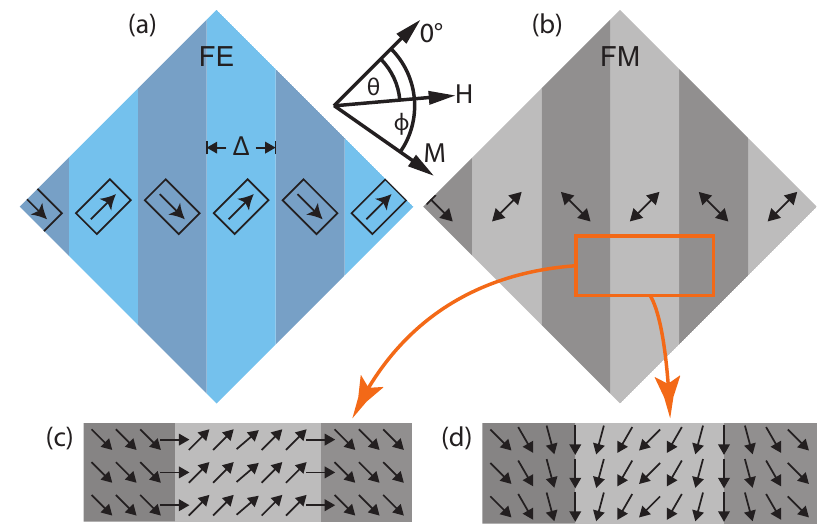}
\caption{\label{Fig1}Schematic illustration of a thin-film multiferroic heterostructure exhibiting domain pattern transfer. The projections of the ferroelectric polarization onto the film plane are indicated by arrows in (a). Magnetoelectric coupling between the two ferroic subsystems induces uniaxial magnetic anisotropies in the ferromagnetic film, which are indicated by double-headed arrows in (b). Because of the lateral modulation of uniaxial anisotropy, magnetically uncharged (c) and charged (d) domain walls form when a magnetic field is applied perpendicular and parallel to the walls, respectively. Panels (c) and (d) illustrate the remanent magnetization state for these two configurations.}
\end{figure} 

The physics of domain pattern transfer in multiferroic heterostructures depends on a competition between the strength of the induced magnetic anisotropy and other relevant energies within the ferromagnetic film. In particular, exchange and magnetostatic interactions oppose the formation of regular magnetic domains.  In this Letter, we report on the scaling of domain pattern transfer in multiferroic heterostructures. Micromagnetic simulations and experiments indicate that the imprinting of ferroelectric domains into a ferromagnetic film varies with domain size and magnetic field direction. Breakdown of pattern transfer occurs when the widths of the ferroelectric domains and magnetic domain walls become comparable. Using a one-dimensional model, we derive concrete expressions for the evolution of the spin rotation between magnetic domains as a function of domain width. Finally, switching between two scaling regimes with different magnetic anisotropy is demonstrated in a rotating magnetic field. The scaling laws can be utilized to optimize ferroelectric-ferromagnetic domain correlations and to design magnetic nanomaterials with field-tunable properties.  

\begin{figure*}
\includegraphics{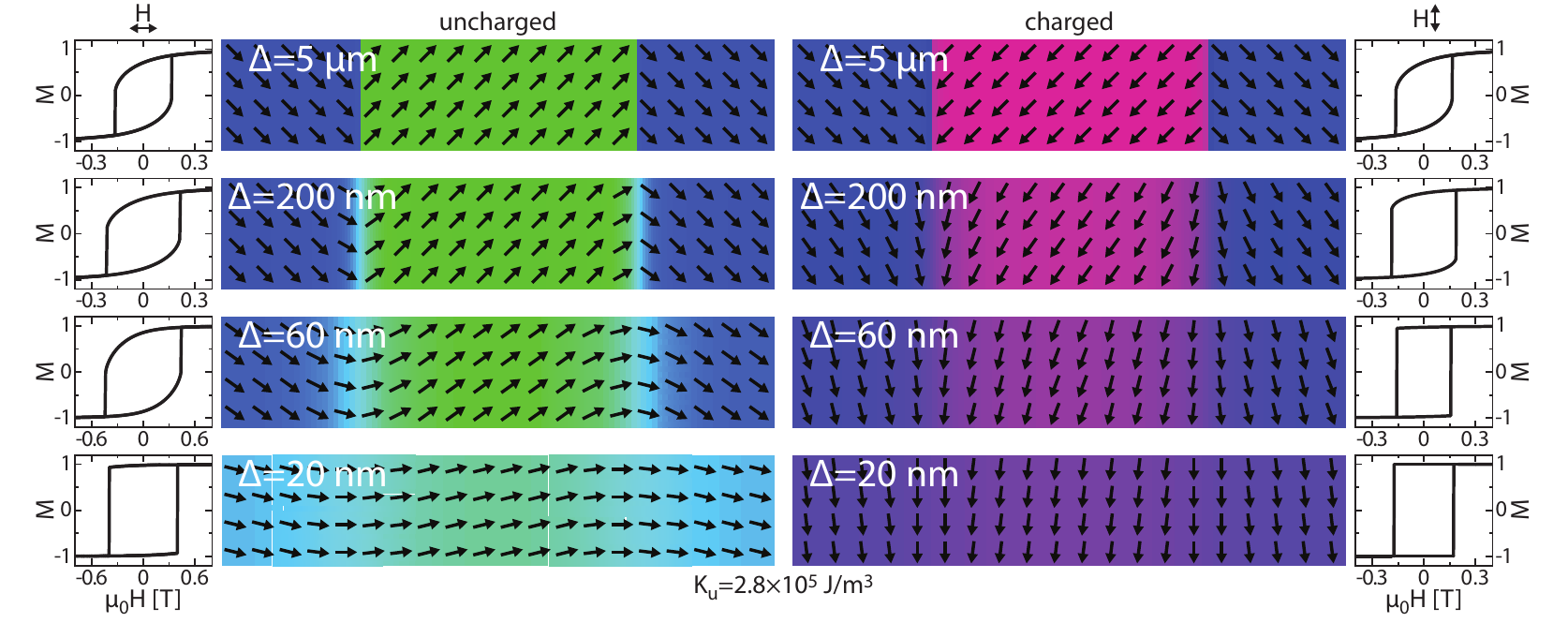}
\caption{\label{Fig2}Simulated remanent magnetization state as a function of domain width for uncharged (left) and charged (right) domain walls, and corresponding hysteresis curves from the whole area for magnetic fields applied perpendicular (left) and parallel (right) to the stripe domains.}
\end{figure*}

The multiferroic heterostructures that we consider consist of a ferromagnetic film that is coupled to a ferroelectric material with regular stripe domains (Fig.~\ref{Fig1}). The in-plane ferroelectric polarization (or its projection) rotates by 90$^\circ$ at domain boundaries, which mimics the geometry of experimentally studied BiFeO$_3$ and BaTiO$_3$ systems \cite{2008NatMa...7..478C,2009PhRvL.103y7601L,2011PhRvL.107u7202H,ADMA:ADMA201100426,2011ITM....47.3768L,2012PhRvB..86a4408C}. Local magnetoelectric coupling via exchange interactions or strain transfer and inverse magnetostriction induces an alternating uniaxial magnetic anisotropy in the ferromagnetic film. Abrupt 90$^\circ$ rotations of the magnetic anisotropy axis strongly pin the magnetic domain walls onto the ferroelectric domain boundaries. As a result, the magnetic domain walls do not move in an applied magnetic field and magnetization reversal proceeds by coherent rotation and abrupt magnetic switching within the domains \cite{ADMA:ADMA201100426,2011ITM....47.3768L}. Because of strong domain wall pinning, two distinctive magnetic microstructures can be initialized in multiferroic systems exhibiting domain pattern transfer \cite{2012PhRvB..85i4423F}: Magnetically uncharged head-to-tail domain walls are stabilized when the magnetic field is applied perpendicular to the stripe domains [Fig.~\ref{Fig1}(c)]. In this case, the intrinsic properties of magnetic domain walls and the coupling between neighboring stripe domains are determined by a competition between short-range ferromagnetic exchange interactions and the strength of the uniaxial magnetic anisotropy. A magnetic field parallel to the domain stripes stabilizes magnetically charged domain walls [Fig.~\ref{Fig1}(d)]. In this configuration, ferromagnetic coupling between domains is dominated by longer range magnetostatic interactions.           

The micromagnetic simulations were conducted using object oriented micromagnetic framework (OOMMF)
software with two-dimensional periodic boundary conditions \cite{NIST,wangtwo-dimensional2010}. In the simulations, the uniaxial magnetic anisotropy that is induced by magnetoelectric coupling ($K_{u}$), the domain width ($\Delta$), and the ferromagnetic film thickness ($t$) were varied. Other parameters included a saturation magnetization of $M_{s}=1.7\times10^6$ A/m, an exchange constant of  $A=2.1\times10^{-11}$ J/m, and an unit mesh of $2\times2\times{t}$ nm. We note that the use of micromagnetic simulations is appropriate for the analysis of domain pattern transfer and magnetization reversal in multiferroic heterostructures \cite{note1}. The magnetic domain wall width in the simulations is defined as      
\begin{equation}
\delta=\int_{-\infty}^{+\infty} cos (\phi\prime)^2 dx,
\end{equation}
where $\phi\prime$ is the reduced magnetization angle 
\begin{equation}
\phi\prime = \left(\phi - \frac{|\phi_{\frac{\Delta}{2}}-\phi_{-\!\frac{\Delta}{2}}|}{2}\right) \cdot \frac{180}{|\phi_{\frac{\Delta}{2}}-\phi_{-\!\frac{\Delta}{2}}|}
\end{equation}
and $\phi_{\frac{\Delta}{2}}$ and $\phi_{-\!\frac{\Delta}{2}}$ are the magnetization angles in the center of two neighboring stripe domains.

\begin{figure*}
\includegraphics{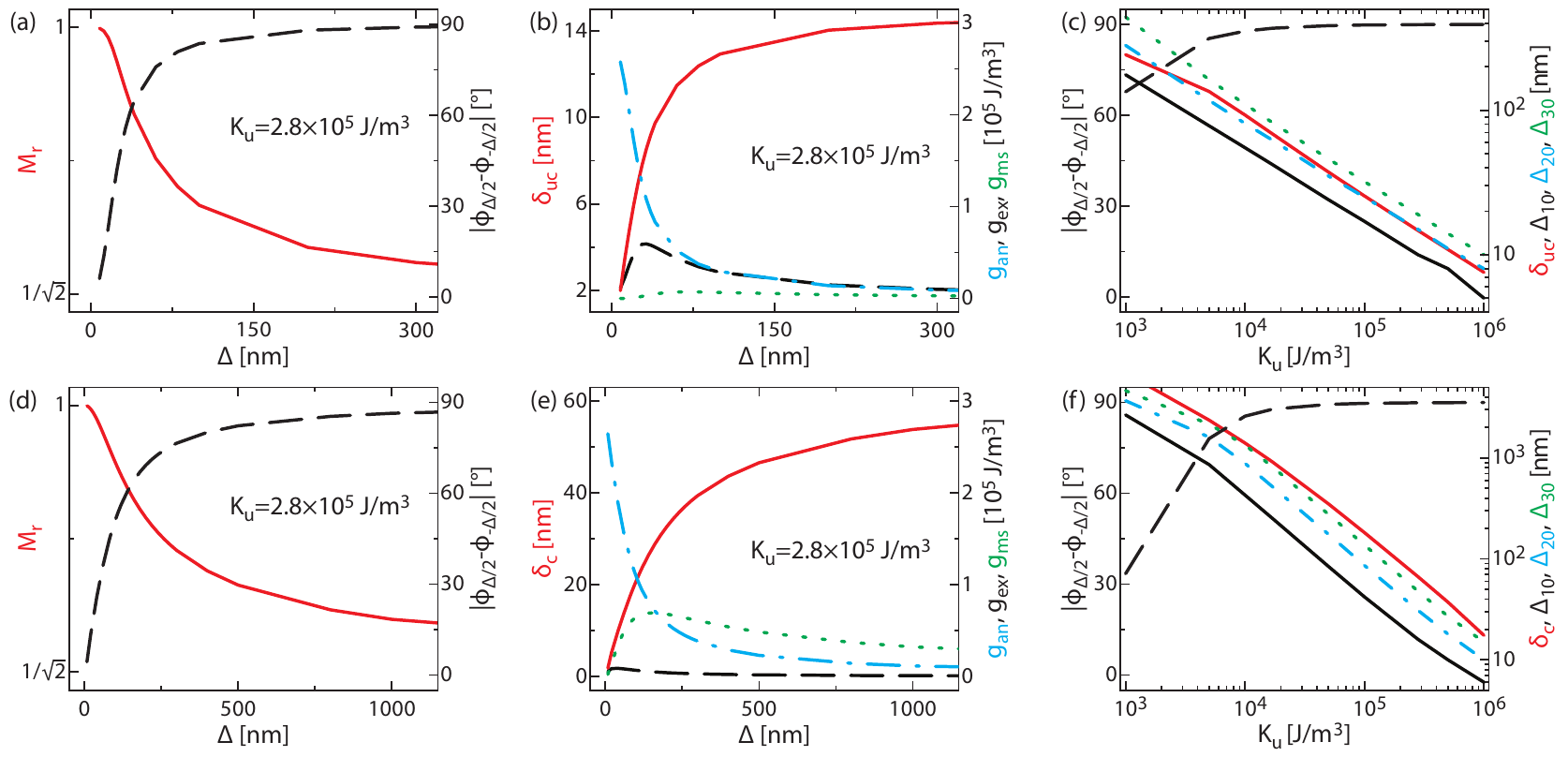}
\caption{\label{Fig3}(a) Film remanence ($M_r$) and spin rotation $|\phi_{\frac{\Delta}{2}}-\phi_{-\!\frac{\Delta}{2}}|$ between domains in the remanent state (dashed line) as a function of domain width ($\Delta$) for uncharged magnetic domain walls. (b) Variation of the domain wall width ($\delta_{uc}$), magnetic anisotropy energy (blue dash-dotted line), exchange energy (black dashed line), and magnetostatic energy (green dotted line) with $\Delta$ for uncharged domain walls. (c) Dependence of the spin rotation between domains and the width of uncharged domain walls on uniaxial magnetic anisotropy. The domain widths for which $|\phi_{\frac{\Delta}{2}}-\phi_{-\!\frac{\Delta}{2}}|$ decrease to \textless 30$^\circ$ ($\Delta_{30}$),  \textless 20$^\circ$ ($\Delta_{20}$), and  \textless 10$^\circ$ ($\Delta_{10}$) are also shown. Panels (d) - (f) contain the same information as (a) - (c) for charged magnetic domain walls. Note that different horizontal scales are used for $\Delta$ to present data for uncharged and charged domain walls.}
\end{figure*}

Figure~\ref{Fig2} shows micromagnetic simulations for uncharged (left) and charged (right) magnetic domain walls as a function of domain width. The images represent the remanent magnetization state for an uniaxial magnetic anisotropy of $K_{u}=2.8\times10^{5}$ J/m$^3$ and a film thickness of $t$ = 10 nm. For a ferroelectric domain width of $\Delta=5$ $\mu$m, the domains are fully imprinted into the ferromagnetic film and the remanent magnetization aligns with the uniaxial easy axes of the magnetoelectric anisotropy, i.e., $|\phi_{\frac{\Delta}{2}}-\phi_{-\!\frac{\Delta}{2}}|=90^\circ$. With reducing domain width, $|\phi_{\frac{\Delta}{2}}-\phi_{-\!\frac{\Delta}{2}}|$ decreases monotonically. As a consequence, the magnetic contrast between domains lowers and the domain pattern vanishes when a nearly ferromagnetic alignment between domains is obtained for small $\Delta$. Breakdown of domain pattern transfer coincides with a change in the magnetization reversal process. For large $\Delta$, the magnetization reverses by coherent spin rotation and abrupt switching within the stripe domains. This is reflected by the rounded shape of the magnetic hysteresis curves with a remanent magnetization of $M_r=1/\sqrt{2}$, which is characteristic of coherent magnetization reversal if the angle between the uniaxial magnetic easy axis and the applied magnetic field is 45$^\circ$ \cite{1948RSPTA.240..599S}. For small $\Delta$, the hysteresis curves become square. In this case, the magnetization aligns with the applied magnetic field and reversal of the field direction triggers an abrupt 180$^\circ$ switch of the nearly uniform film magnetization. The variation of remanent magnetization with $\Delta$ is summarized in Figs. 3(a) and 3(d).  

The data of Fig.~\ref{Fig2} indicate that domain pattern transfer breaks down at a different domain width for uncharged and charged domain walls. Distinctive scaling for these two configurations is explained by dissimilar interaction lengths of the relevant competing energies. Figure~\ref{Fig3} summarizes the evolution of the main magnetic parameters as a function of domain width. For uncharged magnetic domain walls [Figs. 3(a)-(c)], the magnetic system is dominated by the uniaxial anisotropy energy ($g_{an}$) and the exchange energy ($g_{ex}$), whereas the magnetostatic energy  ($g_{ms}$) is small [Fig. 3(b)]. The domain size dependence of this system is therefore characterized by the exchange length, here defined as $l_{ex}=\sqrt{A/K_{u}}$. If $\Delta>>l_{ex}$, the magnetization aligns with the easy anisotropy axes of the individual stripe domains and full pattern transfer is obtained. On the other hand, ferromagnetic exchange interactions force the magnetization of neighboring domains to align parallel when $\Delta<<l_{ex}$. In between these two limiting cases, the spin rotation between domains and the magnetic domain wall width decrease with reducing $\Delta$ [Figs. 3(a) and 3(b)]. Figure 3(c) illustrates the dependence of pattern transfer breakdown on the strength of uniaxial magnetic anisotropy. Plotted are the critical domain widths below which $|\phi_{\frac{\Delta}{2}}-\phi_{-\!\frac{\Delta}{2}}|$ decrease to \textless 30$^\circ$ ($\Delta_{30}$),  \textless 20$^\circ$ ($\Delta_{20}$), and  \textless 10$^\circ$ ($\Delta_{10}$) together with the domain wall width of uncharged domain walls ($\delta_{uc}$) and $|\phi_{\frac{\Delta}{2}}-\phi_{-\!\frac{\Delta}{2}}|$ for 5 $\mu$m wide domains. Clearly, the spin rotation between neighboring magnetic domains becomes small when the width of the ferroelectric domains and the width of magnetic domain walls are comparable ($\Delta_{20}\approx\delta_{uc}$ for the entire anisotropy range). The close agreement between the critical domain width for which pattern transfer breaks down and the width of uncharged magnetic domain walls is explained by their mutual dependence on magnetic anisotropy and ferromagnetic exchange. Both parameters scale linearly with $l_{ex}$, which is confirmed by the 1/$\sqrt{K_{u}}$ dependence of the $\Delta$'s and $\delta_{uc}$ in Fig. 3(c).  

For charged magnetic domain walls [Figs. 3(d)-(f)], the breakdown of domain pattern transfer and the width of the walls are mainly determined by the uniaxial anisotropy energy and the magnetostatic energy [Fig. 3(e)]. As magnetostatic coupling between domains extends over a longer distance than exchange interactions, the magnetization of neighboring domains are forced to align parallel at a considerable larger domain width. Also for this configuration, domain pattern transfer breaks down when the width of stripe domains becomes smaller than the width of magnetic domain walls. The slope of the curves in Fig. 3(f) ($\Delta$'s and $\delta_{c}$) are proportional to 1/$K_u$.

\begin{figure}
\includegraphics{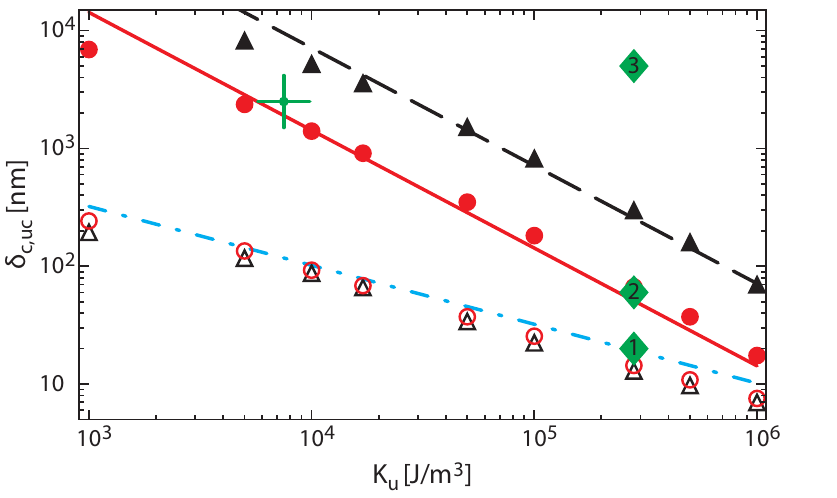}
\caption{\label{Fig4}Simulated uncharged (open symbols) and charged (filled symbols) domain wall width as a function of uniaxial magnetic anisotropy. The circles and triangles indicate data for a magnetic film thickness of 10 and 50 nm, respectively. The lines are calculated using $\delta_{uc}\approx\pi\sqrt{A/2K_{u}}$ and $\delta_{c}\approx\pi\mu_{0}M_{s}^{2}t/8K_{u}$. The green cross corresponds to the parameters of the experimental CoFeB/BaTiO$_3$ sample (Fig. 5).}
\end{figure}

A comprehensive comparison of the breakdown of domain pattern transfer for both magnetic configurations is given in Fig. 4. Here, the width of uncharged and charged domain walls are plotted as a function of uniaxial magnetic anisotropy for a film thickness of 10 and 50 nm. Since the domain wall width is a good measure for the disappearance of magnetic domains, the graph can be considered as a phase diagram: The areas above the curves indicate parameters for which ferroelectric domains are transferred to the adjacent ferromagnetic film, whereas the areas below the curves represent conditions for which the film magnetization is mostly uniform. A good approximation for the width of uncharged domain walls is given by $\delta_{uc}\approx\pi\sqrt{A/2K_{u}}$, which is the width of 90$^\circ$ N\'{e}el walls in the limiting case of zero magnetostatic energy (dash-dotted line in Fig. 4) \cite{Handley}. Although an analytical expression is not available for the width of 90$^\circ$ charged walls, the simulated data of Fig. 4 are in close agreement with $\delta_{c}\approx\pi\mu_{0}M_{s}^{2}t/8K_{u}$, which is half the width of 180$^\circ$ charged domain walls \cite{1979ITM....15.1251H,note3}. 

Because of the two different scaling regimes, it is possible to select parameters between the phase transitions for uncharged and charged domain walls. Under such conditions, rotation of an applied magnetic field would result in the successive writing and erasure of magnetic domains. To verify this prediction, we experimentally studied a 50 nm thick CoFeB film on a BaTiO$_3$ single-crystal substrate with in-plane ferroelectric domains. The strain-induced uniaxial magnetic anisotropy of this film was $K_{u}=7.5\times10^{3}$ J/m$^{3}$, which, according to Fig. 4, positions the parameters of the sample between the phase transitions for uncharged and charged domain walls if 100 nm $<\Delta<$ 10 $\mu$m. Figure 5(a) shows magneto-optical Kerr effect microscopy images of a magnetic stripe domain with $\Delta\approx$ 2.5 $\mu$m. A clear magnetic contrast is obtained when a 8 mT magnetic field is applied perpendicular to the stripe domains, i.e., for uncharged domain walls. Rotation of the magnetic field causing a transition to charged walls ($H$ parallel to stripes) lowers the magnetic contrast and thus the spin rotation between neighboring domains. The polar plot of Fig. 5(b) displays experimental and simulated data on the variation of $|\phi_{\frac{\Delta}{2}}-\phi_{-\!\frac{\Delta}{2}}|$ with the magnetic field direction. Both curves confirm the existence of two scaling regimes for domain pattern transfer and the ability to switch between uniform and patterned magnetic states by magnetic field rotation. 

\begin{figure}
\includegraphics{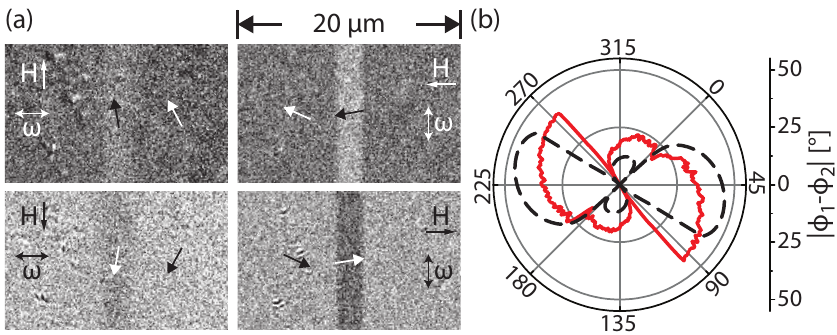}
\caption{\label{Fig5}(a) Magneto-optical Kerr effect microscopy images of magnetic domains in a 50 nm thick CoFeB film on BaTiO$_3$. The images on the left (right) are recorded with a magnetic field of 8 mT parallel (perpendicular) to the stripe domains. The width of the central stripe domain is 2.5 $\mu$m and the uniaxial magnetic anisotropy is $K_{u}=7.5\times10^{3}$ J/m$^{3}$. The sensitive axis of magneto-optical Kerr effect contrast is indicated by $\omega$. (b) Variation of the spin rotation $|\phi_{\frac{\Delta}{2}}-\phi_{-\!\frac{\Delta}{2}}|$ between domains with the direction of a 6 mT magnetic field. The dashed black line and solid red line represent simulated and experimental data, respectively.}
\end{figure}

Finally, we show that the breakdown of domain pattern transfer in multiferroic heterostructures can be calculated using a one-dimensional model when the energy of the system is dominated by magnetic anisotropy and ferromagnetic exchange \cite{porratispatially2002}, i.e., for uncharged domain walls. In this model, two stripe domains of width $\Delta$ are assumed and periodic boundary conditions are imposed. The following analytical expression for the spin rotation between domains is derived by energy minimization:
\begin{equation}
|\phi_{\frac{\Delta}{2}}-\phi_{-\!\frac{\Delta}{2}}|=\phi_{\frac{\Delta}{2}}-\text{acos}\left(\sin\phi_{\frac{\Delta}{2}}\right),
\label{SpinRotEqn}
\end{equation}
where the magnetization angle in the middle of the stripe domains $\phi_{\frac{\Delta}{2}}$ can be calculated as a function of $\Delta$ using:
\begin{equation}
K_u\left(\sin^2\phi_{\frac{\Delta}{2}}\right)
-F\left(\text{asin}\left(\frac{1}{\sqrt{2}\sin\phi_{\frac{\Delta}{2}}}\right)|\sin^2\phi_{\frac{\Delta}{2}}\right)=\frac{\Delta}{2l_{ex}}, \label{Langleprof1}
\end{equation}
and $F(\varphi | m)$ and $K(m)$ are elliptic and complete elliptic integrals of the first kind \cite{abramowitzhandbook1964,gradshteyntable1980}. Results from the one-dimensional model are in excellent agreement with micromagnetic simulations and experimental data.  

In summary, we have shown that domain pattern transfer in multiferroic heterostructures scales with the width of ferroelectric domains. Domain correlations are lost when the ferroelectric domains are narrower than the width of magnetic domain walls. Since two types of magnetic domain walls with different widths can be forced to form in a controlled manner, two scaling regimes are accessible. Recurrent switching between both regimes is possible by rotation of an applied magnetic field. The demonstrated ability to write and erase magnetic domains by magnetic field rotation opens up a vast playground for the design of field-tunable magnetic nanomaterials.  

\begin{acknowledgments}
This work was supported by the Academy of Finland (Grant No. 260361) and the European Research Council (ERC-2012-StG 307502-E-CONTROL). K.J.A.F. acknowledges financial support from the Finnish Doctoral Program in Computational Sciences. The authors thank Tuomas Lahtinen and Arianna Casiraghi for fruitful discussions. 
\end{acknowledgments}


\begin{thebibliography}{10}
\bibitem{note1}%
  \BibitemOpen
  \bibinfo {note} {{Excellent agreements between experiments and micromagnetic
  simulations on CoFe/BaTiO$_3$ have previously been demonstrated in Refs. 11
  and 15.}}%
  \bibAnnoteFile{Stop}{note1}%
\bibitem{note3}%
  \BibitemOpen
  \bibinfo {note} {{The profile of charged domain walls changes when the spin
  rotation is reduced from 180$^\circ$ to 90$^\circ$ [22]. Hence, the
  expression for $\delta_{c}$ can only be used as an approximation for the
  domain wall width.}}%
  \bibAnnoteFile{Stop}{note3}%
\end{thebibliography}
%

\end{document}